\documentclass[aps,pre,twocolumn,groupedaddress]{revtex4}

\def\sR{\hbox{I\kern-.1667em\hbox{R}}}

\def\vn{{\bf n}}

\def\la{\langle}
\def\ra{\rangle}

\def\ep{\epsilon}
\def\trace{\hbox{Tr}}
\def\gn{{\mathcal G}({\bf n})}

\def\Del#1#2{\frac{\partial #1}{\partial #2}}

\def\half{{\textstyle{1\over 2}}}
\def\third{{\textstyle{1\over 3}}}
\def\etal{{\it et al.}}

\begin{document}


\title{Statistical Mechanics and Lorentz Violation}


\author{Don Colladay}
\email[]{colladay@ncf.edu}
\affiliation{New College of Florida}

\author{Patrick McDonald}
\email[]{mcdonald@ncf.edu}
\affiliation{New College of Florida}


\date{\today}

\begin{abstract}
The theory of statistical mechanics is studied in the presence of 
Lorentz-violating background fields.
The analysis is performed using the Standard-Model Extension (SME) together
with a Jaynesian formulation of statistical inference.  
Conventional laws of thermodynamics are obtained in the presence of a perturbed 
hamiltonian that contains the Lorentz violating terms.
As an example, properties of the nonrelativistic ideal gas are calculated in
detail.
To lowest order in Lorentz violation, the scalar thermodynamic variables are only 
corrected by a rotationally invariant combination of parameters that mimics
a (frame dependent) effective mass. 
Spin couplings can induce a temperature independent polarization in the classical gas 
that is not present in the conventional case.
Precision measurements in the residual expectation values of the magnetic moment of 
Fermi gases in the limit of high temperature may provide interesting limits on these 
parameters.

\end{abstract}


\maketitle

\section{Introduction}

The notion that the minimal standard model serves as a low energy
limit to a more fundamental theory which includes a quantum
description of gravity has led to the development of theories which
extend the standard model and predict qualitatively different
physical phenomena \cite{kps}. In this context, a framework for studying the
effects of spontaneous Lorentz symmetry breaking and 
possible resulting CPT violation within the context of conventional
quantum field theory has
been developed \cite{ck,fredef} and studied intensively \cite{cpt98}.  
This framework formally contains all possible operators utilizing
standard model fields that satisfy coordinate reparametrization invariance
and is called the Standard-Model Extension (SME).
The minimal version of the SME is restricted to exhibit a number of useful properties:  
it preserves
energy-momentum conservation, observer Lorentz invariance,
conventional quantization, hermiticity, microcausality, positivity of
the energy, gauge invariance and power counting renormalizability.
In addition, the framework provides a generic model for any theory which extends
the standard model and provides for
spontaneous Lorentz symmetry breaking and CPT violation.
For example, Lorentz violation may arise in noncommutative field
theory \cite{lane} or random dynamics models \cite{nielsen}.

To date, there has been no confirmed evidence for Lorentz violation,
hence it is reasonable to assume that any violation must be small 
in conventional laboratory frames (concordant frames) \cite{ralph}.
Various experiments utilizing mesons \cite{mesons1,mesons2,mesons3}, 
baryons \cite{baryons1,baryons2,baryons3}, electrons \cite{electrons1,electrons2,electrons3},
photons \cite{ph1,ph2,ph3,ph4,ph5}, and muons \cite{muons} have reached a precision that probes
these parameters at Planck-suppressed scales.
Recent analysis have also been extended to the neutrino sector \cite{neutrinos1,neutrinos2},
instantons \cite{instantons}, supersymmetric models \cite{berger},
and the gravitational sector \cite{kgrav}.

Statistical mechanics has been applied to rotationally invariant
CPT-violating terms
in the SME to provide a mechanism for baryogenesis
in thermal equilibrium \cite{bckp},  
but a detailed exposition of the formalism for general Lorentz violation 
is lacking.  
It is the goal of this paper to address this issue.

A general framework is provided in this paper for performing 
statistical mechanics calculations within the context of the
SME.
The required assumptions regarding statistical inference and 
the definitions of relevant thermodynamic quantities are given.
A complete analysis of the effects of all Lorentz-violating
terms on a nonrelativistic ideal gas is performed.
The corresponding
thermodynamics is studied and the various changes in thermodynamic
quantities are analyzed.
In section II, the notation and the relevant formalism
for performing statistical mechanics calculations within the context
of the SME is presented.  Section III contains calculations of the
relevant thermodynamic quantities for single particle systems.
Section IV generalizes the result to a classical gas with a
variable particle number and incorporates the chemical potential.  
Section V generalizes the result to a quantum gas of fermions
and section VI gives corresponding results for spin-0 bosons.  The
conclusions are gathered in section VII.

\section{Notation and Formalism}

Let $\{\psi_i\}_{i=1}^\infty$ be a collection of states and suppose that
 $\{f_j\}_{j=1}^l$ is a finite collection of real valued functions on
 the collection of states. Given a distribution of states, $q_i =
 q(\psi_i),$ we  denote by brackets the corresponding expectations; $\la
 f_j \ra = \sum_i f_j(\psi_i) q_i.$ We denote the (information) entropy
 associated to the distribution $q_i$ by   
\begin{eqnarray}
S & = & -k\sum_i q_i \ln{(q_i)} 
\quad ,
\label{ent1}
\end{eqnarray}
where $k$ is a positive constant that will later be identified with Boltzmann's constant.

Suppose we are given an observation of the mean values $\{\la f_j\ra
\}_{j=1}^l$ and we seek a distribution that best fits our observation.
In his approach to statistical inference, Jaynes \cite{jaynes} argues that one
should choose a distribution which maximizes (\ref{ent1}) subject to
the constraints on expectations given by the observations.  In so
doing, a now standard argument via variational calculus leads
immediately to the solution 
\begin{eqnarray}
q_i & = & \frac{e^{-\sum_{j=1}^l \lambda_j f_j(\psi_i)}}{Z} 
\quad ,
\label{dist1}
\end{eqnarray}
where the Lagrange multipliers, $\lambda_j,$ are real constants and $Z$
is the partition function, $Z({\bf \lambda}) = \sum_{i} e^{-\sum_{j=1}^l
\lambda_j f_j(\psi_i)}.$  From (\ref{dist1}) it follows that 
\begin{eqnarray}
\la f_j\ra & = & - \frac{\partial}{\partial \lambda_j}
\ln(Z) \quad . 
\label{ex2} 
\end{eqnarray}
If, in addition to being functions of state, the $f_j$ are permitted to
depend on a finite collection of parameters, $a_j, \ 1\leq j \leq m,$
then it is easy to see that 
\begin{eqnarray}
\sum_j \lambda_j \left\la \frac{\partial f_j}{\partial a_k} \right \ra & = & -
\frac{\partial}{\partial a_k} 
\ln(Z)
\quad . \label{ex3} 
\end{eqnarray}
The resulting maximal entropy is given by
\begin{equation}
S_{max} = k \ln{Z} + k \sum_j \lambda_j \la f_j \ra
\quad .
\label{entgen}
\end{equation}
Variation of the maximal entropy with respect to the
parameters $\lambda_j$, $\alpha_j$, and $\la f_j \ra$ yields
the relation
\begin{equation}
d S = \sum_k {\partial \ln{Z} \over \partial \alpha_k} d \alpha_k
+ \sum_j \lambda_j d \la f_j \ra 
\quad . 
\label{gfirstlaw}
\end{equation}

As is clear from our development, (\ref{dist1})-(\ref{ex2}) are formal
rules of inference which follow for any system described as above,
under the assumption of maximum entropy.  Jaynes applies these formal
rules  to the study of equilibrium thermodynamics.  More precisely,
suppose that we consider a system of identical particles constrained
to lie in a box of volume $V.$  Suppose that $f_1 = E$ denotes the
energy levels corresponding to the possible states of the system and
that $f_2 = N$ denotes the number of particles in the system.  Given
observations of the mean values, $\la E\ra, \ \la N \ra,$ one obtains
the partition function for the grand canonical ensemble: 
\begin{equation}
Z(\alpha, \beta)  =  \sum_{i,j} e^{ -\beta E_j -\alpha N_i}
\quad ,
\end{equation}
where we have identified the Lagrange multiplier $\lambda_1$ with the
scaled inverse temperature $\beta= \frac{1}{kT}$ and the Lagrange
multiplier $\lambda_2 = \alpha$ with $ -\beta \mu,$ where $\mu$ is the
chemical potential.   

Allowing the energy to depend on the volume $V$ and
setting $a_1=V$ in (\ref{ex3}) we get the usual description of
pressure 
\begin{equation}
P = 
- \left\la \frac{\partial E}{\partial V} \right \ra  =  
\frac{1}{\beta} \frac{\partial}{\partial V} 
\ln(Z)
\quad . \label{ex4} 
\end{equation}
Equations (\ref{ent1})-(\ref{ex4}) give the central
features of thermodynamics including expressions for the entropy as a
function of temperature, volume and particle number, and
the usual first law of thermodynamics.  The standard
expression for average energy and particle number are given by
(\ref{ex2}) and the expressions for specific heat at constant volume
and at constant pressure follow by taking the appropriate derivatives
of the entropy:
\begin{eqnarray}
C_V & = & T\left(\Del{ S}{T}\right)_{\la N \ra, V} = \left(\Del{\la
E\ra}{T}\right)_{\la N \ra,V} \quad ,
\label{sph1}\\
C_P & = & T\left(\Del{ S}{T}\right)_{\la N \ra, P} \label{sph2} 
\quad .
\end{eqnarray}
Thermodynamic potentials such as the Helmholtz free energy are defined
in the usual way.

As was emphasized by Jaynes, the method of maximum entropy provides
accurate predictions of thermodynamic properties, assuming empirically
accurate mean values and laws of motion embodied in the associated
hamiltonian.  For the purpose of probing the thermodynamics of systems
of particles with symmetry violations, the method offers a simple 
framework for generating results.  All that remains is to specify the
relevant underlying hamiltonian.  

In the case of free
spin-$\frac12$ Dirac fermions $\psi$ of mass $m$, the SME is determined
by the lagrangian \cite{ck}
\begin{equation}
{\cal{L}} = \half i\bar{\psi} \Gamma_\nu  \stackrel{\leftrightarrow} {\partial^\nu} \psi -
\bar{\psi} M \psi 
\quad ,
\label{l1}
\end{equation}
where 
\begin{eqnarray*}
\Gamma_\nu & = &   \gamma_\nu + c_{\mu\nu}
\gamma^\mu + d_{\mu\nu}\gamma_5\gamma^\mu + e_\nu + i f_\nu \gamma_5 +
\half g_{\lambda \mu \nu}\sigma^{\lambda \mu} ,\\
M & = & m + a_\mu \gamma^\mu + b_\mu \gamma_5\gamma^\mu + \half
H_{\mu\nu} \sigma^{\mu \nu} \quad .
\end{eqnarray*}
In the above expressions $a_\mu, \ b_\mu, \ c_{\mu \nu }, \ d_{\mu \nu }, \ e_\mu, \ f_\mu
, \ g_{\lambda \mu \nu}$ and $H_{\mu \nu}$ are real fixed background parameters which
determine the Lorentz violation.  For low energy applications the
associated nonrelativistic hamiltonian $H$ has been obtained using
a generalized Foldy-Wouthuysen transformation \cite{klane}.
To second order in
${\bf p}/{m}$,  this hamiltonian is given by\cite{fn1}
\begin{equation}
H = \frac{p^2}{2m} + H_{LV} \quad ,
\label{H1}
\end{equation}
with
\begin{eqnarray}
H_{LV} & = & A + B_j\sigma^j + C_{j}\frac{p_j}{m} +
D_{jk}\frac{p_j}{m} \sigma^k + \nonumber \\ 
 &  &  F_{jk}\frac{p_jp_k}{2m} +
G_{jkl}\frac{p_jp_k}{2m} \sigma^l \quad , 
\label{H2}
\end{eqnarray}
where common terms in the original lagrangian (\ref{l1}) are collected:
\begin{equation}
A = (a_0-mc_{00} - me_0) \quad ,
\end{equation}
\begin{equation}
 B_j = (-b_j + md_{j0} -
\half m \ep_{jkl}g_{kl0} + \half \ep_{jkl}H_{kl})
\quad ,
\end{equation}
\begin{equation}
C_j = [a_j - m(c_{0j} + c_{j0}) - m e_j]
\quad ,
\end{equation}
\begin{eqnarray}
D_{jk} =& & \left[-b_{0}\delta_{jk} + m(d_{kj}+ d_{00}\delta_{jk})
\right. \nonumber \\
& & \left. + m\ep_{klm}(\half g_{mlj} + g_{m00}\delta_{jl}) +
\ep_{jkl}H_{l0}\right]  , 
\end{eqnarray}
\begin{equation}
F_{jk} = - 2 \left[(c_{jk} +
\half c_{00}\delta_{jk})\right]
\quad ,
\end{equation}
\begin{eqnarray}
G_{jkl} & = & 2 \left\{\left[(d_{0j} + d_{j0}) \right. \right.\nonumber \\
& - & \left. \left.\half (b_j/m + d_{j0} +
\half \ep_{jmn}(g_{mn0} + 
H_{mn}/m))\right]\delta_{kl}\right. \nonumber \\ 
 & + & \left. \half (b_l/m
+ \half \ep_{lmn}g_{mn0})\delta_{jk} \right. \nonumber \\
& - & \left. \ep_{jlm}(g_{m0k} +
g_{mk0})\right\} \quad .  
\end{eqnarray}
The hamiltonian is invariant under spatial translations, therefore
the logarithm of the associated grand partition function scales with volume
as in the conventional case.
Using Eq.\ (\ref{ex4}) together with this property yields the 
relation 
\begin{eqnarray}
\frac{PV}{kT} & = &  \ln(Z)
\quad .
\label{PVZ}
\end{eqnarray}

\section{Single particle systems}

We first consider a system consisting of a single free spin-$\frac12$ 
particle governed by the hamiltonian $H$ appearing in (\ref{H1}), constrained
to a cube of side length $L$.  For small violations of Lorentz
symmetry, we derive the corresponding perturbations to the statistical
mechanics using the single particle partition function.
This simplifies the initial analysis by eliminating the need to discuss
the chemical potential.

The standard unperturbed solutions are written in the form
\begin{eqnarray}
\psi^{(0)}_{\vn, s}(\bf x) & = & \prod_i \sin{\left(\frac{n_i\pi x_i}{L}\right)}
\chi_s 
\quad ,
\label{es1} 
\end{eqnarray}
where $\vn = (n_1,n_2,n_3)$ is a triple of positive integers and 
$s \in \{1,-1\}$ denotes a sign. 
Note that the two-component spinor $ \chi_s$ may depend on $\bf n$.
The corresponding unperturbed energy levels are written as
\begin{equation}
E_{\vn,s}^{(0)} = \frac{\pi^2\hbar^2}{2mL^2} n^2
\quad . 
\end{equation}  

The first order correction to the energy levels due
to the Lorentz-violating terms are found using standard
degenerate perturbation theory as:
\begin{equation}
\langle \psi_{\vn,s}| H_{LV} |  \psi_{\vn,s} \rangle
= {\pi^2 \hbar^2 \over 2 m L^2} (A n^2 + \sum_i F_{ii}n_i^2 
+ s |\gn|),
\label{EA1}
\end{equation}
where the vector $\gn$ is defined with components
\begin{equation}
(\gn)_j \equiv {2 m L^2 \over \pi^2 \hbar^2}B_j + \sum_i G_{iij}n_i^2
\quad .
\end{equation}
The $\vn$ dependent spinors $\chi_s$ have been chosen to satisfy
$\gn \cdot {\bf \sigma} \chi_s = s \chi_s$ and therefore diagonalize 
$H_{LV}$.
Note that $C$ and $D$ perturbations depend
linearly on momentum and therefore don't contribute
a correction to the energies. 

The perturbed energy is written as 
\begin{eqnarray}
E_{\vn, s} & = & E^{(0)}_{\vn,s} +\delta E_{\vn,s}
\quad ,
\label{Eperturb} 
\end{eqnarray}
with $\delta E_{\vn,s}$ given by the matrix elements
(\ref{EA1}).  The partition function for a single particle
becomes 
\begin{eqnarray}
Z^{(1)} & = & \sum_\vn e^{-\beta E_{\vn,+}} + e^{-\beta E_{\vn,-}}
\quad .
\label{Z2}
\end{eqnarray}
Approximating the sum on the right hand side of (\ref{Z2}) by the
appropriate integrals, we obtain expressions for partition functions
corresponding to hamiltonians with Lorentz-violating terms.  
The partition function is even in the spin-dependent correction
terms and therefore there are no lowest order corrections contributed by $B$ or $G$. 
The $A$ and $F$ terms correct the partition function as 
\begin{eqnarray}
Z^{(1)}  & \simeq & 2 e^{-\beta A} \frac{4\pi}{8} \int_0^\infty
n^2 e^{-\beta \frac{\pi^2 \hbar^2}{ 2mL^2} (n^2 + \sum F_{ii}n_i^2)} dn \nonumber \\ 
& \simeq & 2e^{-\beta A}n_Q V (1 - \half \trace(F))
\quad ,
\label{ZA}
\end{eqnarray}
where $V$ is the volume of the box,  $n_Q = 
({m}/{2\beta \pi \hbar^2})^{{3}/{2}}$ is the
quantum concentration, and
$\trace(F) = \sum F_{ii}.$  

Using the relationship $\la E^{(1)} \ra =
- (\partial / \partial \beta) 
\ln(Z^{(1)})$ (cf (\ref{ex2})),  it follows from (\ref{ZA}) that
only the $A$ term corrects the energy. 
The $A$ term corresponds to a constant shift in all of the energy
levels, hence it is possible to redefine the zero point of the
energy \cite{fn2} to eliminate this
contribution.

The correction to the partition function due to the $F$ term can 
be incorporated into an effective mass for the fermion 
\begin{equation}
m^* = (1 - \third \trace(F)) m
\quad ,
\label{effmass}
\end{equation}
because only the rotationally invariant trace component contributes.
The $\trace(F)$ term appears to be a trivial scaling that can be absorbed
into the mass, but this is not quite true because
the effective mass violates boost invariance.
This means that the effective mass can be different in various
laboratory frames.
However, such effects will be relativistically suppressed and 
therefore difficult to observe using Earth-based experiments. 

More interesting is the correction to the expectation value of the
spin.
In this case, only the $B$ and $G$ terms will contribute.
The expectation is calculated as
\begin{eqnarray}
\la {\bf s}^{(1)} \ra & = &(Z^{(1)})^{-1} \sum_{\vn, s}
\frac{s \gn}{|\gn|} e^{-\beta\frac{\pi^2\hbar^2}{2mL^2}(n^2 + s |\gn|)} 
\nonumber \\
 & \simeq &- (n_Q V)^{-1} \beta \frac{\pi^2\hbar^2}{2mL^2}\sum_{\vn}
\gn e^{- \beta \frac{\pi^2\hbar^2}{2mL^2}n^2} \nonumber \\ 
 & \simeq & - \beta {\bf B} - \frac{1}{2} \trace(\bf G) 
 \quad ,
\label{SP1G} 
\end{eqnarray}
where the vector $(\trace({\bf G}))_k \equiv \sum_i G_{iik}$ is defined.  
Note that while it is not in general possible to simultaneously diagonalize the
contribution from $B$ and $G$ terms, the same computation allows us to
treat the case where both terms occur.  The corresponding spin
expectations decouple.  

\section{Classical gas}

We now introduce the chemical potential and consider a classical gas of 
free spin-$\frac12$ Dirac Fermions.  
As in section II, $N$ denotes the (variable) particle
number for the system.  
The grand partition function for the
system can be written in terms of the single-particle partition
function as
\begin{eqnarray}
Z^{(C)} & = & \exp\left( e^{-\alpha^{(C)}} Z^{(1)}(\beta) \right)
\quad ,
\label{gcepart} 
\end{eqnarray}
where $\alpha^{(C)} = -\beta \mu^{(C)}$ and $\mu^{(C)}$ is the chemical potential
of the classical gas.  
Thus equation (\ref{ZA}) also gives the first order corrections 
for the grand partition function.  Using
the grand partition function, the resulting
expressions for expected particle number and energy are
\begin{eqnarray}
\la N^{(C)} \ra & = & e^{-\alpha} Z^{(1)} \quad , \label{n31} \\
\la E^{(C)} \ra & = &  \frac{3}{2} \la N^{(C)} \ra kT \quad .  \label{e31}
\end{eqnarray}
Since $PV = kT \ln{Z^{(C)}},$ it follows that there is no change in the 
classical ideal gas law, even in the presence of Lorentz violation.

Solving for the chemical
potential $\mu^{(C)}$ (to lowest order in Lorentz violating parameters) yields
\begin{eqnarray}
\mu^{(C)} & = & -kT\left( \ln\left(\frac{2n_Q}{n^{(C)}}\right) -
\half\trace( F) \right) 
\quad ,
\label{chem1} 
\end{eqnarray}
where $n_Q$ is the quantum concentration and $n^{(C)} \equiv {\la
N^{(C)} \ra}/{V}$ is the concentration of the classical gas.  
We conclude that in the presence of Lorentz violation there is a change in the chemical
potential as expected from the effective mass argument presented in
the previous section.    

The corresponding entropy $S^{(C)}$ is found using equation (\ref{entgen}) as
\begin{eqnarray}
S^{(C)} & = & \alpha^{(C)} \la N^{(C)} \ra k + \frac{5}{2} \la N^{(C)}
 \ra k
 \quad .
 \label{s31}
\end{eqnarray}
Using (\ref{chem1}), the modified Sackur-Tetrode equation is
\begin{eqnarray}
S^{(C)} & = &  \la N^{(C)} \ra k\left[ \frac{5}{2} -
\frac 1 2 \trace (F) +  \ln\left(\frac{2n_Q}{n^{(C)}}\right)
\right]
~ . 
\label{ST1} 
\end{eqnarray}
From (\ref{ST1}) and (\ref{sph1})-(\ref{sph2}) it is clear that there
is no change to the specific heat.  

Finally, we verify that the expectation of the spin is in fact the
single particle result in equation (\ref{SP1G}) times the
expected particle number 
\begin{equation}
\la {\bf s}^{(C)} \ra = - \la N^{(C)} \ra [ \beta {\bf B} 
+ \half \trace(\bf G) ]
\quad .
\end{equation}

\section{Quantum Gas - Fermions}

Next, the quantum occupancy of the orbitals is restricted to
be $0$ or $1$ and the low-temperature limit is analyzed.
With notation from previous sections, the partition function for the grand
canonical ensemble associated to a Fermi gas is
\begin{eqnarray}
Z^{(Q)}(\alpha) & = & \prod_{\vn, s} \left(1 +
e^{-\alpha}e^{-\beta E_{\vn, s}}\right)
\quad .
\label{ZFerm1}  
\end{eqnarray}
In the unperturbed case, the grand partition function is
calculated using the appropriate integral approximation as
(zero subscripts represent unperturbed quantities)
\begin{eqnarray}
\ln\left(Z^{(Q)}_0(\alpha_0)\right) & = & \pi \int_0^\infty
n^2 \ln \left( 1+ e^{-\alpha_0 -\beta \frac{\pi^2 \hbar^2}{2mL^2} n^2}
\right) dn \nonumber \\
 & = & \frac{2}{\lambda^3} V f_{\frac{5}{2}}(e^{-\alpha_0}) 
 \quad ,
 \label{Qintegral}
\end{eqnarray}
where $\lambda = {h}/{(2\pi mkT)^{\frac{1}{2}}}$ is the thermal
wavelength (related to the quantum concentration by $1/\lambda^3 = n_Q$)
and $f_{\nu}(e^{-\alpha})$ is the Fermi-Dirac integral   
\begin{eqnarray}
f_{\nu}(e^{-\alpha}) & = & \frac{1}{\Gamma(\nu)} \int_0^\infty
\frac{x^{\nu-1}}{e^{\alpha} e^x +1} dx \quad .
\label{Fint1}
\end{eqnarray}
Using (\ref{ex2}) gives the standard results
\begin{eqnarray}
\la N^{(Q)}_0(\alpha_0) \ra & = & \frac{2}{\lambda^3}  V
f_{\frac{3}{2}}(e^{-\alpha_0}) \quad ,\label{QN1}\\ 
\la E^{(Q)}_0(\alpha_0) \ra & = &  \frac{3}{2} \la N^{(Q)}_0
(\alpha_0)\ra kT
\frac{f_{\frac{5}{2}}(e^{-\alpha_0})}{f_{\frac{3}{2}}(e^{-\alpha_0})}
\label{QE1}   
\quad ,
\end{eqnarray}
and the quantum ideal gas law 
\begin{eqnarray}
\frac{P}{n k T} & = &
\frac{f_{\frac{5}{2}}(e^{-\alpha_0})}{f_{\frac{3}{2}}(e^{-\alpha_0})}
\quad .
\label{IGL1} 
\end{eqnarray}

As in the classical case, there are no first order corrections to the
partition function for Lorentz violating terms given by the
coefficients $B, \ C, \ D$ and $G$ appearing in (\ref{H2}).  For
violating terms of type $F,$ a change of variables in the integral
(\ref{Qintegral}) gives first order corrections for the partition
function which we write as 
\begin{eqnarray}
\ln\left(Z^{(Q)}(\alpha)\right) & \simeq & \left( 1 - \half
 \trace(F)\right) \ln\left(Z^{(Q)}_0(\alpha) \right)~. \label{QZ2}
\end{eqnarray}
As in the classical case, only the rotationally invariant component
of $F$ contributes, therefore it is possible to absorb the
term into an effective mass as before.
Standard calculations then immediately give the results
\begin{eqnarray}
\la N^{(Q)}(\alpha) \ra & = &  \left( 1 -  \half
\trace(F)\right) \frac{2}{\lambda^3} f_{\frac{3}{2}}(e^{-\alpha})
\quad ,
\label{QN2}\\ 
\la E^{(Q)}(\alpha) \ra & = &   \frac{3}{2} \la N^{(Q)}(\alpha)
\ra kT\frac{f_{\frac{5}{2}}(e^{-\alpha})}{
f_{\frac{3}{2}}(e^{-\alpha})}\quad ,
\label{QE2} 
\end{eqnarray}
as well as the ideal gas law
\begin{eqnarray}
\frac{P}{n^{(Q)} k T} & = &
\frac{f_{\frac{5}{2}}(e^{-\alpha})}{f_{\frac{3}{2}}(e^{-\alpha})}
\quad ,
\label{IGL2} 
\end{eqnarray}
where $n^{(Q)}$ is the concentration of the quantum gas 
(not to be confused with the quantum concentration $n_Q$ defined
earlier).
Note that the ideal gas law is in fact modified in the quantum
case due to its dependence on $\alpha \ne \alpha_0$.
The correction due to $F$ may be incorporated easily by
replacing $\lambda(m)$ by $\lambda(m^*)$ using the
effective mass given in equation (\ref{effmass}).

From (\ref{Fint1}) it is clear that the map $\alpha \to
f_{\frac{3}{2}}(e^{-\alpha}) $ is strictly monotonic and thus
invertible.  Writing the inverse as ${\mathcal F},$ we have a
formal expression for the chemical potential
\begin{eqnarray}
\mu^{(Q)} & \simeq- k T {\mathcal F}\left(\frac{\lambda^3(m^*)
n^{(Q)}}{2}\right) \quad .
\label{chemQ1}
\end{eqnarray}

Equation (\ref{chemQ1}) demonstrates that conventional formulas
can be used to obtain the relevant quantities.
These formulas can be found in a standard statistical mechanics 
text such as \cite{Pa}.
For instance, the chemical potential at zero temperature 
defines the Fermi energy and is modified as
\begin{equation}
\mu^{(Q)}(T=0) \equiv {\cal{E_F}} \simeq {{\cal{E_F}}^{(0)}} (1 + \third \trace(F))
\quad ,
\label{chem32}
\end{equation}
where ${{\cal{E_F}}^{(0)}} = (\hbar^2 / 2 m)(3 \pi^2 n^{(Q)})^{2/3}$
is the conventional Fermi energy.
Approximating to the next highest order in temperature gives
\begin{eqnarray}
\mu^{(Q)} & \simeq & {\cal{E_F}} \left[ 1-\frac{\pi^2}{12}
\left(\frac{kT}{{\cal E_F}}\right)^2\right] . \label{chem33}
\end{eqnarray}
Using (\ref{QN2}), (\ref{QE2}), an asymptotic expansion for
$f_{\frac{5}{2}}$ and (\ref{chem33}) we obtain
\begin{eqnarray}
\frac{\la E^{(Q)} \ra}{\la N^{(Q)} \ra} & \simeq & \frac{3}{5} {\cal 
E_F}\left[ 1 + \frac{5\pi^2}{12} \left(\frac{kT}{{\cal
E_F}}\right)^2\right] . \label{en2} 
\end{eqnarray}
From (\ref{sph1}) and (\ref{en2}) we get an expression for the
specific heat in the limit of low temperature
\begin{equation}
\frac{C_V}{\la N^{(Q)} \ra k}  \simeq  \frac{\pi^2}{2} \frac{kT}{{\cal 
E_F}} \simeq \frac{\pi^2}{2} \frac{kT}{{\cal 
E_F}^{(0)}} (1 - \third \trace(F))
\quad .
\end{equation}
Writing the entropy using (\ref{entgen}) gives the
perturbed entropy in the low temperature limit as
\begin{equation}
S^{(Q)}  \simeq  C_V \quad , 
\end{equation}
as expected from the integrated equation for specific heat.
The low-temperature ideal gas law perturbation is
\begin{equation}
P \simeq \frac{2}{5}n^{(Q)} {\cal{E_F}} \left[1 
+ {5 \pi^2 \over 12}\left( {kT \over {\cal{E_F}}}\right)^2 \right]
\quad .
\label{SQ22}
\end{equation}

The expectation value for the spin can be calculated in the quantum
regime using the fractional occupancies 

\begin{equation}
f(\vn,s)  =  \frac{1}
{e^{\alpha} e^{\beta E_{\vn,s}}+1} \quad ,
\label{fol}
\end{equation}
where $E_{\vn, s} = E^{(0)}_{\vn, s} + \delta E_{\vn, s}$ 
and the perturbation is as given in (\ref{EA1}).  With notation
as above, the expectation of the spin is given by
\begin{eqnarray}
\la {\bf s}^{(Q)} \ra & = & \sum_{\vn, s} s \frac{\gn}{|\gn|}
f(\vn, s) \nonumber \\
 &\simeq & -2\beta \frac{\pi^2 \hbar^2}{2mL^2} \sum_{\vn} \gn
\frac{e^{\alpha} e^{\beta \frac{\pi^2
\hbar^2}{2mL^2}n^2}}{(1+e^{\alpha} e^{\beta \frac{\pi^2
\hbar^2}{2mL^2}n^2})^2} \nonumber \\
& \simeq & -\la N^{(Q)} \ra 
\left[2 {\beta  \over \lambda^3}f_{\half}(e^{-\alpha}) {\bf B}
+ {1 \over 2} \trace({\bf G})\right] .
 \label{spinG}
\end{eqnarray}
This calculation demonstrates the surprising fact that the 
contribution of the $G$ term is 
temperature independent and does not randomize at high temperature.
At low temperatures, the contribution from the $B$ term
can be written as
\begin{equation}
\la {\bf s}^{(Q)}_B \ra \simeq -  \la N^{(Q)} \ra 
\frac 3 2 {{\bf B}\over {\cal{E_F}}}
\left[ 1 - {\pi^2 \over 12}\left( {kT \over {\cal{E_F}}}\right)^2 \right] 
\quad .
\end{equation}
This is identical to the result for a quantum Fermi gas in an
external magnetic field when only spin interactions are relevant.
Note that the parameter ${\bf B}$ is a fixed background vector that does
not rotate with the experiment.

\section{Quantum gas - Bosons}

It is possible to generate a model for a free spin-0 boson gas by combining
two fermions into a singlet representation of the spin group.
This means that the expectation value of all spin-couplings
vanish in the hamiltonian (\ref{H2}).
The resulting hamiltonian is given by 
\begin{eqnarray}
H & = & \frac{p^2}{2m} + A +  C_{j}\frac{p_j}{m} +
F_{jk}\frac{p_jp_k}{2m} 
\quad .
\label{H3}
\end{eqnarray}
Choosing the ground state energy to be zero and employing the notation of the
previous sections, the associated grand partition function for the
unperturbed case is 
\begin{eqnarray}
\ln(Z^{(QB)}_0(\alpha_0)) & = & -\sum_\vn \ln(1-e^{-\alpha_0}e^{-\beta
E^{(0)}_{\vn}}) \nonumber \\
& & - \ln(1-e^{-\alpha_0}) \quad ,
\end{eqnarray}
where $E^{(0)}_{\vn} = (\pi^2 \hbar^2/{2mL^2})n^2$ as before,
and the ground state has been separated out to allow for
Bose-Einstein condensation at low temperatures. 
Approximating the sum as an integral gives 
the standard result
\begin{equation}
\ln(Z^{(QB)}(\alpha_0)) =  \frac{1}{\lambda^3}Vg_{\frac{5}{2}} (e^{-\alpha_0}) -
\ln(1-e^{-\alpha_0})  
\quad ,
\end{equation}
where $\lambda $ is the thermal wavelength and 
$g_{\frac{5}{2}}(e^{-\alpha})$ is the Bose-Einstein 
integral   
\begin{eqnarray}
g_{\nu}(e^{-\alpha}) & = & \frac{1}{\Gamma(\nu)} \int_0^\infty
\frac{x^{\nu-1}}{e^{\alpha} e^x -1} dx
\quad .
\label{gint1}
\end{eqnarray}
Using (\ref{ex2}) we obtain the expected number of particles in
the excited states and the associated energy as
\begin{eqnarray}
\la N^{(QB)}_0(\alpha_0) \ra - \la N_{G0} \ra & = & \frac{1}{\lambda^3}  V
g_{\frac{3}{2}}(e^{-\alpha_0}) \quad ,\label{BN1}\\ 
\la E^{(Q)}_0(\alpha_0) \ra & = &  \frac{3}{2}  kT \frac{V}{\lambda^3}
g_{\frac{5}{2}}(e^{-\alpha_0}) \quad ,
\label{BE1}   
\end{eqnarray}
where $\la N_{G0} \ra = [e^{\alpha_0} - 1]^{-1}$ is the expected
number of particles condensed into the ground state.  

The only nontrivial leading order perturbation in (\ref{H3}) arises from the $F$
term.
A calculation which follows that done for the case of fermions gives 
\begin{eqnarray}
\ln\left(Z^{(QB)}(\alpha)\right) & \simeq & \left( 1 - \half
 \trace(F)\right) \frac{1}{\lambda^3}Vg_{\frac{5}{2}} (e^{-\alpha}) \nonumber \\
 & & -
\ln(1-e^{-\alpha}) \quad . 
\label{QZB2}
\end{eqnarray}
It follows from (\ref{ex2}) and (\ref{QN1})-(\ref{QZ2}) that for the
perturbed case we have 
\begin{eqnarray}
\la N^{(QB)}(\alpha) \ra - \la N_{G0} \ra & = &  \left( 1 -  \half
\trace(F)\right) \frac{1}{\lambda^3} g_{\frac{3}{2}}(e^{-\alpha}) ~~,
\nonumber \\ 
\la E^{(QB)}(\alpha) \ra & = &   \frac{3}{2} kT \frac{V}{\lambda^3}
g_{\frac{5}{2}}(e^{-\alpha}) \quad .
\label{QEB2}  
\end{eqnarray}
As in the Fermi case, the chemical potential can be expressed 
as a function of the number of particles in excited states.  
The computations, modulo
obvious modifications, are similar to (\ref{chemQ1})-(\ref{SQ22}). 
Because only $\trace(F)$ enters into the grand partition function,
it is possible to use the concept of effective mass to absorb the
effect as before.
Standard results of Bose-Einstein condensation therefore hold
in a given laboratory frame.
The effect is nontrivial, as in the fermion case, because
boosting the experiment will change the effective mass.
Modifications of these calculations are required if the ground
state wave function does not exhibit rotational symmetry, such as is
the case in actual condensate experiments due to some
optical and magnetic trapping configurations \cite{becond}.  
In this case, the zero point of energy can depend on orientation
of the apparatus and induce rotational variations in the condensate
properties.
An analysis of this type for specific experiments would be interesting, 
but is beyond the scope of the paper.

\section{Conclusion}

The formalism of statistical mechanics in the presence of symmetry violation
parallels the conventional 
situation.  The laws of thermodynamics are the same as in the
conventional case, although the specific expectation values of thermodynamic 
quantities can be
modified by the Lorentz-violating terms.  
Our approach involves an
assumption of a maximal lack of information (or entropy) regarding a system
subject to various constraints imposed by the physical observables.
The temperature and chemical potentials can be defined simply
as the lagrange multipliers associated with the constraints, hence
two systems in equilibrium automatically have the same
temperature and chemical potential as there is only one lagrange
multiplier for each overall constraint.
This method in fact produces equivalent results to the more conventional assumption
of equal a-priori statistics,
however, it has the advantage of providing straightforward definitions of 
thermodynamic variables with conventional equilibrium properties.

As an explicit example, the unperturbed system was assumed to be an ideal gas 
in the absence of any external applied fields (such as magnetic or gravitational fields).  
Expectation values for scalar thermodynamic quantities such as energy and particle
number were unaltered except for an overall scaling factor $\trace (F)$.
This happens because the unperturbed system is rotationally invariant
and scalar expectation values can only be corrected by perturbations with
commensurate symmetry. 
This can also be incorporated as an effective mass
$m^* = (1 - \third \trace (F)) m$ in the hamiltonian, although
the effective mass defined in this way depends on the observer's Lorentz frame.
For instance, a gas in motion on the surface of the Earth would exhibit 
slight sidereal variations in effective mass due to changes in $\trace (F)$.
This effect is relativistically suppressed and unlikely to be observable for
physically reasonable values for the violation parameters.

More interesting is the net spin expectation value contributed by the terms
that couple to the spin.  The pure-spin coupling $B_j$ mimics a constant background magnetic
field and induces a corresponding magnetic moment per unit volume in the gas. 
Any additional applied magnetic field could be added to this term to
calculate the net result on an actual experiment.
For example,
an Earth-based experiment will rotate in space
and the constant background vector will
interfere with the applied magnetic field to produce a sidereal variation.

The derivative-spin coupling $G_{ijk}$ generates a fundamentally new type of effect
that induces a temperature-independent polarization in the classical gas
that is proportional to $\trace({\bf G})$.
This means that even at very high temperatures there will be a net residual polarization 
that does not randomize.  
This is a reasonable result, considering that the $\trace({\bf G})$ term couples the spin to the 
conventional kinetic energy term in the hamiltonian and indicates that any physical effects 
should scale accordingly.
An effect of this type should be clearly distinguishable from stray magnetic
field effects that have an inverse temperature dependence.

In the zero temperature limit, both the $B_j$ and $G_{ijk}$ terms contribute
a polarization to the Fermi gas.  
The ${\bf B}$ contribution depends on $n^{1/3}$
while the $\trace({\bf G})$ depends linearly on particle density $n$.
This means that the effects of the $\trace({\bf G})$ term can grow significantly
faster with increasing density than conventional magnetic field effects.


\begin{thebibliography}{MM1}

\bibitem{kps}
V.A.\ Kosteleck\'y and S.\ Samuel,
Phys.\ Rev.\ D {\bf 39}, 683 (1989);
{\it ibid.} 
{\bf 40}, 1886 (1989);
Phys.\ Rev.\ Lett.\ {\bf 63}, 224 (1989);
{\it ibid.} 
{\bf 66}, 1811 (1991);
V.A.\ Kosteleck\'y and R.\ Potting,
Nucl.\ Phys.\ B {\bf 359}, 545 (1991);
Phys.\ Lett.\ B {\bf 381}, 89 (1996);
Phys.\ Rev.\ D {\bf 63}, 046007 (2001); 
V.A.\ Kosteleck\'y, M.\ Perry, and R.\ Potting,
Phys.\ Rev.\ Lett.\ {\bf 84}, 4541 (2000). 

\bibitem{ck} 
D.\ Colladay and V.A.\ Kosteleck\'y,
Phys.\ Rev.\ D {\bf 55}, 6760 (1997);
Phys.\ Rev.\ D {\bf 58}, 116002 (1998).

\bibitem{fredef}
D.\ Colladay and P.\ McDonald,
J.\ Math.\ Phys.\ {\bf 43} 3554 (2002).

\bibitem{cpt98}
For a summary of recent theoretical models and
experimental tests
see, for example,
{\it CPT and Lorentz Symmetry}, V.A.\ Kosteleck\'y, ed., 
World Scientific, Singapore, 1999; 
{\it CPT and Lorentz Symmetry II}, V.A.\ Kosteleck\'y, ed.,
World Scientific, Singapore, 2002.

\bibitem{lane}
S.\ M.\ Carroll, {\it et al.}, Phys.\ Rev.\ Lett.\ 
{\bf 87} 141601 (2001);
Z. Guralnik, {\it et al.}, Phys.\ Lett.\ B
{\bf 517} 450 (2001). 

\bibitem{nielsen}
C.\ D.\ Froggatt and H.\ B.\ Nielsen,
hep-ph/0211106.

\bibitem{ralph}
V.\ A.\ Kosteleck\'y and R. Lehnert,
Phys.\ Rev.\ D {\bf 63}, 065008 (2001).

\bibitem{mesons1}
KTeV Collaboration,
H.\ Nguyen, in Ref.\ \cite{cpt98};
OPAL Collaboration,
R.\ Ackerstaff \etal,
Z.\ Phys.\ C {\bf 76}, 401 (1997);
DELPHI Collaboration,
M.\ Feindt \etal,
preprint DELPHI 97-98 CONF 80 (1997);
BELLE Collaboration,
K.\ Abe \etal,
Phys.\ Rev.\ Lett.\ {\bf 86}, 3228 (2001);
BaBar Collaboration,
B.\ Aubert
\etal, 
hep-ex/0303043;
FOCUS Collaboration,
J.M.\ Link \etal, 
Phys.\ Lett.\ B {\bf 556}, 7 (2003).

\bibitem{mesons2}
D.\ Colladay and V.A.\ Kosteleck\'y,
Phys.\ Lett.\ B {\bf 344}, 259 (1995);
Phys.\ Rev.\ D {\bf 52}, 6224 (1995);
Phys.\ Lett.\ B {\bf 511}, 209 (2001);
V.A.\ Kosteleck\'y and R.\ Van Kooten,
Phys.\ Rev.\ D {\bf 54}, 5585 (1996);
O.\ Bertolami \etal,
Phys.\ Lett.\ B {\bf 395}, 178 (1997);
N.\ Isgur \etal,
Phys.\ Lett.\ B {\bf 515}, 333 (2001).

\bibitem{mesons3}
V.A.\ Kosteleck\'y,
Phys.\ Rev.\ Lett.\ {\bf 80}, 1818 (1998);
Phys.\ Rev.\ D {\bf 61}, 016002 (2000);
Phys.\ Rev.\ D {\bf 64}, 076001 (2001).

\bibitem{baryons1}
L.R.\ Hunter \etal,
in 
V.A.\ Kosteleck\'y, ed.,
\it CPT and Lorentz Symmetry, \rm
World Scientific, Singapore, 1999;
D.\ Bear \etal,
Phys.\ Rev.\ Lett.\ {\bf 85}, 5038 (2000);
D.F.\ Phillips \etal,
Phys.\ Rev.\ D {\bf 63}, 111101 (2001);
M.A.\ Humphrey \etal,
Phys.\ Rev.\ A {\bf 68}, 063807 (2003);
Phys.\ Rev.\ A {\bf 62}, 063405 (2000);
V.A.\ Kosteleck\'y and C.D.\ Lane,
Phys.\ Rev.\ D {\bf 60}, 116010 (1999).

\bibitem{baryons2}
R.\ Bluhm \etal,
Phys.\ Rev.\ Lett.\ {\bf 88}, 090801 (2002);
Phys.\ Rev.\ D {\bf 68}, 125008 (2003).

\bibitem{baryons3}
F.\ Can\`e \etal,
physics/0309070. 

\bibitem{electrons1}
H.\ Dehmelt \etal,
Phys.\ Rev.\ Lett.\ {\bf 83}, 4694 (1999);
R.\ Mittleman \etal,
Phys.\ Rev.\ Lett.\ {\bf 83}, 2116 (1999);
G.\ Gabrielse \etal,
Phys.\ Rev.\ Lett.\ {\bf 82}, 3198 (1999);
R.\ Bluhm \etal,
Phys.\ Rev.\ Lett.\ {\bf 82}, 2254 (1999);
Phys.\ Rev.\ Lett.\ {\bf 79}, 1432 (1997);
Phys.\ Rev.\ D {\bf 57}, 3932 (1998).

\bibitem{electrons2}
B.\ Heckel,
in Ref.\ \cite{cpt98};
L.-S.\ Hou, W.-T.\ Ni, and Y.-C.M.\ Li,
Phys.\ Rev.\ Lett.\ {\bf 90}, 201101 (2003);
R.\ Bluhm and V.A.\ Kosteleck\'y,
Phys.\ Rev.\ Lett.\ {\bf 84}, 1381 (2000).

\bibitem{electrons3}
H.\ M\"uller, S.\ Herrmann, A.\ Saenz,
A.\ Peters, and C.\ L\"ammerzahl,
Phys. Rev. D {\bf 68}, 116006 (2003).

\bibitem{ph1}
S.M.\ Carroll, G.B.\ Field, and R.\ Jackiw, 
Phys. Rev. D {\bf 41}, 1231 (1990);
M.P.\ Haugan and T.F.\ Kauffmann,
Phys. Rev. D {\bf 52}, 3168 (1995);
V.A.\ Kosteleck\'y and M.\ Mewes,
Phys.\ Rev.\ Lett.\ {\bf 87}, 251304 (2001).

\bibitem{ph2}
R.\ Jackiw and V.A.\ Kosteleck\'y,
Phys.\ Rev.\ Lett.\ {\bf 82}, 3572 (1999);
C.\ Adam and F.R.\ Klinkhamer,
Nucl.\ Phys.\ B {\bf 657}, 214 (2003);
H.\ M\"uller, C.\ Braxmaier, S.\ Herrmann, 
A.\ Peters, and C.\ L\"ammerzahl,
Phys. Rev. D {\bf 67}, 056006 (2003);
T.\ Jacobson, S.\ Liberati, and D.\ Mattingly,
Phys.\ Rev.\ D {\bf 67}, 124011 (2003);
V.A.\ Kosteleck\'y and A.G.M.\ Pickering,
Phys.\ Rev.\ Lett.\ {\bf 91}, 031801 (2003);
R.\ Lehnert, 
Phys.\ Rev.\ D {\bf 68}, 085003 (2003);
G.M.\ Shore, 
Contemp.\ Phys.\ {\bf 44}, 503 {2003};
B.\ Altschul, 
Phys.\ Rev.\ D {\bf 69}, 125009 (2004);
hep-th/0407172.

\bibitem{ph3}
V.\ A.\ Kosteleck\'y, M.\ Perry,
and R. Lehnert, 
Phys. Rev. D {\bf 68}, 123511 (2003).

\bibitem{ph4}
J.\ Lipa \etal,
Phys.\ Rev.\ Lett.\ {\bf 90}, 060403 (2003);
H.\ M\"uller \etal,
Phys.\ Rev.\ Lett.\ {\bf 91}, 020401 (2003);
P.\ Wolf \etal,
gr-qc/0401017.

\bibitem{ph5}
V.A.\ Kosteleck\'y and M.\ Mewes,
Phys.\ Rev.\ D {\bf 66}, 056005 (2002);
V.A.\ Kosteleck\'y and Q. Bailey,
hep-ph/0407252.

\bibitem{muons}
V.W.\ Hughes \etal,
Phys.\ Rev.\ Lett.\ {\bf 87}, 111804 (2001);
R.\ Bluhm \etal,
Phys.\ Rev.\ Lett.\ {\bf 84}, 1098 (2000).

\bibitem{neutrinos1}
S.\ Coleman and S.L.\ Glashow, 
Phys.\ Rev.\ D {\bf 59}, 116008 (1999);
V.\ Barger, S.\ Pakvasa, T.J.\ Weiler, and K.\ Whisnant, 
Phys.\ Rev.\ Lett.\ {\bf 85}, 5055 (2000);
J.N.\ Bahcall, V.\ Barger, and D.\ Marfatia,
Phys.\ Lett.\ B {\bf 534}, 114 (2002);
V.A.\ Kosteleck\'y and M.\ Mewes, hep-ph/0308300;
S.\ Choubey and S.F.\ King, 
Phys.\ Lett.\ B {\bf 586}, 353 (2004).

\bibitem{neutrinos2}
V.A.\ Kosteleck\'y and M.\ Mewes, 
Phys.\ Rev.\ D {\bf 69}, 016005 (2004);
hep-ph/0406255.

\bibitem{instantons}
D. Colladay and P. McDonald,
J.\ Math.\ Phys.\ {\bf 45}, 3228 (2004).

\bibitem{berger}
M.\ S.\ Berger and V.A.\ Kosteleck\'y,
Phys.\ Rev.\ D {\bf 65}, 091701 (2002);
M.\ S.\ Berger,
Phys.\ Rev.\ D {\bf 68}, 115005 (2003).

\bibitem{kgrav}
V.\ A.\ Kosteleck\'y,
Phys.\ Rev.\ D {\bf 69}, 105009 (2004).

\bibitem{bckp}
O.Bertolami, {\it et al.},
Phys.\ Lett.\ B {\bf 395}, 178 (1997). 

\bibitem{jaynes}
E. T. Jaynes,
Phys.\ Rev.\ {\bf 106}, 620 (1957).

\bibitem{klane}
V.A.\ Kosteleck\'y and C.D.\ Lane,
J.\ Math.\ Phys.\ {\bf 40}, 6245 (1999).

\bibitem{fn1}
Note that the components of $\bf p$ are written as $p_j$ in this
paper and the nonrelativistic metric is taken to be $\delta_{jk}$. 
These components correspond to the components $p^j$ defined in
reference \cite{klane}.  This accounts for the sign changes that occurr
in the odd momentum terms in the hamiltonian. 

\bibitem{fn2}
The $A$ term can make important contributions to the physics in the relativistic
limit where energy splittings between particles and antiparticles become relevant,
as is the case in the baryogenesis calculations performed in reference \cite{bckp}.

\bibitem{Pa}
R.\ K.\ Pathria,
{\it Statistical Mechanics}, Oxford Press, Boston, 1996.

\bibitem{becond}
For a review of Bose-Einstein condensation see, for example,
F. Dalfovo, {\it et al.},
Rev.\ Mod.\ Phys.\ {\bf 71}, 463 (1999).

\end{thebibliography}
\end{document}